# Homeland Defense and Security Universal Interface Software (HDUIS) Protocol Communication Gateway UIS Protocol Enhancements, Alterations and Attachments

Dr. Carol A. Niznik
NWSYSTEMS
Rochester, NY
dr_carol_niznik@yahoo.com

**ABSTRACT -** ***The Universal Interface Software(UIS) Protocol was a Theater Missile Defense Gateway Protocol which linked the Strategic Defense Initiative(SDI) Architecture Killer Satellite Software Protocol to the National Test Bed Simulation Software Protocol to enable neural network shock loop operation when ICBMS were approaching the SDI Shield. A Gateway Software is required for Homeland Defense and Security Systems to communicate the sensor information from hardware and software boxes at airports and government buildings and other locations to the Global Information Grid(GIG). Therefore, a Homeland Defense and Security UIS(HDSUIS) Protocol is achieved by UIS conversion to HDSUIS for Thresholds Stabilization and GIG and terrorist sensor Enhancements, Homeland Defense and Security Lagrangian equation and GIG simulation facility timing chart Alterations, and two Catastrophe Theory Protocol Attachments to the UIS Geometric software structure inner cube. This UIS Protocol conversion to the HDSUIS Protocol will track and provide a Congestion Controlled, i.e.,prevention of deadlock and livelock, communication of (1) Shoe bombers and copycat shoe bombers, (2) deeply buried and imbedded boxes with explosives, (3) damage to lase1 equipment, (4) shoulder missile fired armament, and (5) surface to air missiles from their sensor equipment to the Global Information Grid with Theater Missile Defense Characteristics. The Homeland Defense and Security GNNO(Geometric Neural Network Overlay) Protocol will be derived as a conversion of the UIS GNNO Protocol.***

***Keywords-Global Information Grid(GJG); SDI Slield; Catastrophe Theory Protocol; shoe bomber; GIG Simulation Facility; sensor equipment; Strategic Defense Initiative(SDI); Lagrangian equation; UIS Protocol; HDSUIS Protocol; Homeland; Theater Missile Defense; GNNO Protocol***

## I. INTRODUCTION

The Enhancements, Attachments and Alterations of the Universal Interface Protocol are summarized here in reference to the sections of the Universal Interface Software[17] that must be changed to enforce the creation of the Homeland Defense and Securit; Universal Interface Software(HDSUIS) Protocol and it software code to be imbedded within microcircuits within the sensor circuitry. The Enhancement Conversions to the following UIS Protocol sections and partitions with the Catastrophe.theory equations.and critical point explosive detectiOn are: (l) Optimal Subnetwork for Load Percentage, (2) Optimal Load % for N Subnetworks, (3) Contention Resolution Network Load Optimization, (4) CTPA Protocol Theoretical Overlay parameters, and (5) CCPDP Protocol Geometric Protocol Structure. The Enhancement Conversions realize added Threshold Stabilization beyond game theory to the UIS Protocol. The Alteration Conversions will be communication to the Global Information Grid(GIG) software simulation facility realizing a timing chart, which optimally performs network communication to the Homeland Defense and Security sensors, instead of the National Test Bed Simulation facility, the Theater Missile Defense analogy to the Homeland Defense and Security GIG. The Attachment Conversions will be the Catastrophe theory equations for sensing and then communication of the identified image according to the Catastrophe equation's shape identification and correlated Critical Point sensing for explosive detection to the GIG, to determine actions to be taken near or at the sensors to prevent terrorism.

## II. UIS TO HDSUIS ENHANCEMENT CONVERSION: THRESHOLD STABILIZATION FROM NTB TO GIG

The Homeland Defense Global Grid Simulation Facility and the National Test Bed Simulation Facility both utilize the software theory of Threshold Stabilization[20] and the theory of games,. which is analogous to the classical theory of von Neuman's minimax theorem. When two stable entities a and b are separated by a threshold S, are in competition on a domain U, the Threshold Stabilization will be characterized by the evolution of almost all of U toward S and only a fragment of the domain being free to oscillate between a and b. The state of operation is as if all conflict evolves to minimize the damage that results. The damage can be interpreteted in dynamical terms, as the total density of local Catastrophes of the domain. This type of evolution will be dominated by an overall development of the forms of those having the fewest Catastrophes, i.e. the least complex and thus the most stable with a local entropy increase.

The Theater Missile Defense(TMD) National Test Bed(NTB) Simulation facility analogy to the Homeland Defense and Security Global Grid Simulation facility realizing the threshold stabilization and theory of games[24] occurs in the Mid-Course-End Game Phases mideling in the National Test Bed. A critical time interval sequence occurs during the Midcourse Phase handover period[20]. During this time interval, the command and control(C2) section transmits final commands to the weapons battle manager(BM) to execute the end game. This timing can be analytical to timing events transmitted from the DARPA Beacons to the Global Grid for shoe bombers, copycat shoe bombers, deeply buried boxes and laser equipment damage due to explosives and their timing activation at different locations near or in the DARPA Beacons. In the transition to the end-game operational phase, the weapon has been command guilded to the target location by the Battle Manager. Therefore, the weapon is under positive control of the BM/C2 elements. The end game denotes the operational state whereby the SDI weapon, the Kinetic Kill Vehicle, operates exclusively as a homing interceptor. The end game(threshold stabilization) will result in a Catastrophe[24] condition, and not be enabled if congestion control is not integrated into the Strategic Defense Initiative(SDI) Communication System. Therefore, the delays referenced are an example of theoretical upper and lower bound delays computed to aid in the prevention of deadlock and livelock for the comparison to real delays required in the Congestion Controlled Sequential Contention Resolution(CCSCR) Protocol in the Killer Satellite Protocol routing decisions.

## III. THRESHOLD STABILIZATION

Threshold Stabilization**(TS)** is defined by the following eight tuple of characteristics:

$$TS = < L, S, Sg, Fm, G, To, B, Lr, > \quad (1)$$

**Loop g** defined by internal parameters (x,y) with a spatial significance. **Stratification** S, due to a transformation and the map (x,y) ⟶(u,x) is defined by u = $Vx^x$ v = $Vy^x$ **Singularities s**, whose codimensions are too large appear in a stable way, **Formal mechanism Fm** Dynamic **(M,X)** undergoes bifurcation in the sense of Hopf. **Theory of Games, G** (1) Overall evolution of the forms to those having the fewest catastrophes are the least complex and most stable with an increase in local entropy.(2) Classical Theorem Of the Theory Of Games von Neuman's Minimax Theorem states when two stable regimes a and b separated by a threshold S, are in competition on a domain U, the threshold stabilization will be characterized by the evolution of almost all of U toward S with only a fragment of the domain remaining free to oscillate between a and b. The process will behave as if two players a and b compete against each other at each point of U, each adopting the common strategy minimizing their losses (3)Anthropomorphically all conflict evolves so as to minimize the damages that results, (4) Damage- In dynamical terms the total density of local catastrophes of the domain, (5) Evolution dominated by overall evolution of forms to those having the fewest Catastrophes, resulting in increase in local entropy, **Tools,To** correspond to a smoothing a threshold stabilization between these phases(two phases). **Inverse Hyperbolic Breaker B** (1) Beginning stable cusp transforms to a hollow wave, (2)Threshold Stabilization, Critical Moment- representing transition between elliptical and hyperbolic, Parabolic State – can vary locally or either side of the threshold, (3) Parabolic Umbilic.With its critical set C has a double point at the origin with distinct tangents. (4) In image space **O** u,v for w > 0 each branch of the critical curve_has two cusps. **Localization and Reversibility of Transitions Lr** Functional Catastrophe Variation of a growth wave F: $R^3$ x T → U, Universal Unfolding of Catastrophe Local organizing center **O** corresponds to a stratum S of the bifurcation set $F^1(0)$ gives the critical point,

## IV. UIS TO HDSUIS ALTERATION CONVERSION: GIG SIMULATION FACILITY TIMING CHART

For the HDSUIS Protocol the Intelligence Equation is required to establish Threshold Stabilization to prevent extreme Catastrophe and will identify the exact timing of events in a section, where a clock establishes the main time of hardware transitted during the transmission to the GIG. Therefore, the max-min Lagrangian Equation predicts the events and identiifies the timing of Terrorist events as part of the timing chart similar to the Design Automation simulation of the Mother board and Daughter cards for large main frame computers like the IBM 360/65 and 370/85 Computers. The printed circuit board(PCS) runs checked wiring of the Daughter cards(many identical in a unit) and its logic changes onto the Mother Board, realizing a Design Automation simulation. The UIS Protocol Cerebellum Geometric Process and Transit Optimization, i.e., the overall Lagrangian optimization equation for the geometric forms structure, in conjunction with its constraints, is stated as the objective function of effectiveness of computational storage and time complexity intelligence levels in this section of the human brain. The Lagrangian Deformable tensor can be incorporated in the constraint equation to represent motion of terrorists prior to their igniting explosives. The multicriterion Lagrangian optimization equation realizing the biochemical and computer networking constraints is stated in the following equation. The Optimal Intelligence Equation representing the amount of optimal delay for each type of terrorist activity follows,

$$\text{IDE} = \text{AntiTerrorist Objective Function} + a[A - CA] + \beta[B - CB] + \delta[D - CD] + \phi[G - CG] + \gamma[P - CP] + \eta[N - CN] + k[K - CK] \quad (2)$$

**where, $a, \beta, \delta, \phi, \gamma, \eta, k$ are the respective Lagrange Multipliers for the following constraints,**

**CA = Shoe Bomber Constraint,**
**CB = Copycat Shoe Bomber Constraint,**
**CD = Deeply imbedded boxes Constraint,**
**CG = Laser damage airbourne Constraint,**

**CP = Laser damage ground Constraint,**
**CN = Shoulder Missile Firing Armament Constraint**
**CK = Surface to Air Missile Constraint,**

Critical points for this equation will be derived by the CCPDP Protocol if selected from within the HDSUIS Protocol. The HDSUIS Protocol Software will create the optimal type of terrorist delay timing information to be transmitted to the GIG for simulation operations

## V. UIS TO HDSUIS ATTACHMENT CONVERSION: CTPA GATEWAY SOFTWARE FUNCTION

The CTPA Protocol is essentially a Catastrophe Theory Gateway software protocol transmitting terrorist activity in software partitions to the Global Grid, which return with information to enable correction of the damaged softwiJ,fe. The CTPA Protocol contains as the focal Protocol section the GNNO(Geometric Neural Network Overlay) Protocol[l8], because SDIO(Strategic Defense Initiative Organization) Overlay Protocols contained the mathematics to performance evaluate the SDIO and other software for total damage. Therefore, the software can appear to be transfigured, i.e. changed in an outward format, to appear corrected, when special changes for security purposes are being communicated from the Global Grid. In this way, the terrorists are being outsmarted, while this transfiguration enabled by the CTPA Protocol Gateway software allows the time to make the necessary security changes enabled by the Global Grid.

Singular events mathematically characterize events that occur as a result of terrorist actions. Here, the CTPA Protocol, an antiterrorist software attachment protocol enabling communication with the Global Grid is theoretically derived and demonstrated on both an Electronic Battlefield(WTPPP) Protocol[10,11] and a Theater Missile Defense(MICR) Protocol[l5]. The respective application of the CTPA Protocol to the WTPPP Protocol and the MICR Protocol, and the CTPA Software Attachment mechanisms in the CTPA Protocol and their geometric enabling structures is observed. The theory of forecasting disasters for the GIG is also derived.

## VI. CATASTROPHE THEORY FORMALISM

Catastrophes are defined as significant changes arising as a sudden response of a system to a smooth change in external conditions. The geometric interpretation of Catastrophe theory[23,24] is the pattern formed by a single equation of a curve in the plane. Solid analytic geometry represents the transition to a surface. The four dimensional geometry of relativity theory represents a three dimensional hypersurface with two equations for a 2-surface and three for a curve. This process is the

**VII.**

definition of codimension of the object, i.e., the number of equations for a geometric object is equal to the difference between the dimension of the object and that of the space in which it is embedded. In the case where equilibrium states of a process form a surface of some dimension in this space, the projection of the equilibrium surface onto the plane of control parameters can have geometric singularities, which then predict the geometry of the Catastrophe. Therefore, the definition of the Equilibrium state of a system, i.e. the stability domain, which does not bifurcate, is the basis of the catastrophe theory. The seven elementary catastrophes are the definitions; fold, cusp, swallow tail, elliptic umbilic, hyperbolic umbilic, butterfly, and parabolic umbilic. The Catastrophe **C** is mathematically formalized as [24], the eight tuple of Manifold Equilibrium Surfaces,

$$C = < f, c, st, eu, bu, b, pu, ww > \quad (3)$$

**f = fold,** $3x^2+u = 0$ (4)

**c = cusp,** $4x^3+2ux+v = 0$ (5)

**st = swallowtail,** $5x^4 + 3ux^2+2vx + w = 0$ (6)

**eu = elliptic umbillic,** $x^2 - y^2+2wx - u = 0$ (7)

$- 2xy + 2wy + v = 0$ (8)

**hu = hyperbolic umbillic,** $3x^2 + wy - u = 0$ (9)

$3y^2 + wx - v = 0$ (10)

**b=butterfly,** $6x^5 + 4tx^3 + 3ux^2 + 2vx + w = 0$ (11)

**pu =parabolic umbilic** $2xy + 2wx - u = 0$ (12)

$x^2 + 4y^3 2ty - v = 0$ (13)

**ww = wigwam,** $x^7 + sx^5 + tx^4 + ux^3 + vx^2 + wx$ (14)

## CIPA PROTOCOL GEOMETRIC SOFIWARE STRUCTURES AND HARDWARE IMPLEMENTATlON

The CTPA Protocol is represented by the octagonal GNNO Attachment Protocol in the center and the four pentagons representing the eight catastrophe theory geometric structures. Two of the triangles above and below the four pentagons mathematically are the eight catastrophes. The third triangle above and below each of the four pentagons is the geometric structure, where the Max-Min Lagrangian Capacity and delay optimization occurs for the software partitions that have been damaged by the catastrophes in the pentagon containing that triangle.

Since the GNNO Protocol is the octagonal structure that connects the four pentagons characterizing and determining the other catastrophe software partition losses, it can communicate to the four triangles in the four pentagonal structures other Lagrangian optimization parameters. In the MICR Protocol[15] a form of Constraint Removal Lagrangian optimization tracked various failures during transmission and launching of the ICBM and the Kinetic Kill Vehicle. A formulation of this tracking Constraint Removal Lagrangian optimizatuion also is mathematically formulated in this Communication Parity Triangle of each of the pentagonal structures of the CTPA Protocol. The definition of parity realizes the equivalence, i.e. similarity, that is correlated to the Intelligence Distribution release identifying weapons of mass destruction with delay as a constraint in the Lagrangian Optimization in the Communication Parity Triangle.

The SDIO software procedures prevent damage of software required using the Universal Interface Software(UIS) Gateway Software Protocol and the correlated communicating SDIO Ground Simulation Network, i.e., the NTB Centralized Core And Distributed Inner Core (CCDIC) Protocol[20] and the SDIO {\rchitecture Killer Satellite Protocol, i.e., the Congestion Controlled Sequential Contention Resolution (CCSCR) Protocol[8]. For weapons of mass destruction, the SDIO maintained a total containment attitutude for the percentage of Intelligence release[15]. Both Congestion Controlled Gateway Protocol software theory and Two Point Boundary Value Problem equations[51 and their wavelet formulations[19] are required for the mathematical modeling of percentage of intelligence release by the Global Grid for interception of weapons of mass destruction. The CTPA Attachment of the MICR Protocol[15] is a Geometric Square structure. represents The CTPA Attachment of the WTPPP Protocol[10,11] is represented as a Hexagon Software Structure.

## VIII. CIPA GNNO STRATEGIC DEFENSE PLATFORM

The GNNO Protocol[18] forms an overlay for the Universal Interface(UIS) Protocol with the Data Glove realizing that the GNNO Protocol and the UIS Protocol have identical geometric software structures. The mathematical basis for the geometric software structure of both the GNNO Protocol and the UIS Protocol is the Category Theory Principle by Pawlak that there are two sections to the geometric software structure, i.e., the Admissibility and the Construction Sections. Category Theory was used by M. Novotny to formally represent the simulation of a network or system via the derivation of the optimal number of simulation loops required to represent the exact or a bounded number of the geometric structure processes defined by Petri Net processes. The UIS Protocol and the GNNO Protocol geometric softwares are composed of two cubes, the outer cube for the Construction Property and the inner cube for the Admissibility Properties. The CTPA GNNO Protocol is represented by the shaded outer attachments to the CTPA octagon, where the two four sided sections of its geometric Lagrangian equation content are the four planes of the Outer Construction Cube Communication Parity Objective Function, and its overlay linkage to the Inner Admissibility Cube for the four constraints of two capacity groupings, tracking, weapons of mass destruction and n capacity groupings.

This outer attachment of the CTPA GNNO Protocol Strategic Defense Platform for Global Grid Transmission is illustrated by opening up the two CTPA GNNO Cubes and attaching them to the two sides of the bottom of the CTPA Octagon. Thus, the Construction Outer Cube four Communication Parity objective function sections are attached to one end of the four sections of the bottom of the CTPA octagon and the Admissibility Cube four sections of constraints are attached to the other four sides of the bottom of the CTPA octagon represented by the outer octagon shaded sections. This outer Platform section surrounds the CTPA Protocol Octagon for hardware attachment. The inner eight crosshatched shaded sections form an additional Platform Corridor for hardware attachment to the CTPA Protocol, whose area links the CTPA GNNO Protocol Admissibility Cube constraints and the Construction Cube objective function in the Lagrangian optimization equation.

## IX. CTPA PROTOCOL COMMUNICATION PARITY GEOMETRIC SOFTWARE STRUCTURE

The number of equilibrium points, also defined in Catastrophe Theory[24], represent the number of exercises in a War Game[10], which characterize the specific amount of Intelligence Distribution exposure, and are the solution points of the two point boundary value problem. The two dimensional Intelligence Distribution requiring the wavelet formulations of Two Point Boundary Value Problem[5,6] partial differential equations to indicate the most accurate solution for these points. The three nine way Exclusive Or circuits are the CTPA Protocol hardware required to communicate to the Global Grid. The CTPA Protocol in the eight triangles of the four pentagons of the various seven elementary catastrophes and the wigwam dual catastrophe and the dual and the symbolic versions of the eight catastrophes is the conceptual mathematical reason for the CTPA three nine way Exclusive Or circuits hardware implementation.

## X. UIS TO HDSUIS ATTACHMENT CONVERSION: CCPDP PROTOCOL OVERVIEW

The Catastrophic Critical Point Detection Prediction(CCPDP) Software Protocol theoretically determines eight Catastophe Theory equations and adaptation Catastrophe Theory equation critical points from sensor data mathematically modeled manifold angular perturbations and correlated Lagrangian delay equation critical points to enable sensor prediction of terrorist attempts to create catastrophic events. The detection by sensors for Theater Missile Defense and Homeland Defense and Security will be enhanced by the CCPDP Software shape intelligence. The theoretical development of the Catastrophe Theory Critical Point Detection is for Lagrangian delay equations and the eight Catastrophe theory equations; and catastrophe theory adaptation equation critical points.

## XI. CCPDP PROTOCOL GEOMETRIC SOFTWARE STRUCTURES

The CCPDP Protocol geometric structure is composed of three layers of geometric star structures and a rectangular structure overlay. Network intrusion detection is achieved by the prediction of critical points in various levels of computer communication networks for both homeland security sensing of shoe bombers, copy cat shoe bombers and hard and deeply buried targets containing explosives and Theater Missile Defense Airborne Laser component damage. The mathematical formalism of Catastrophe theory[24] develops the CCPDP Software Protocol to ensure AI security in equipment maintenance.

The CCPDP Software Protocol implementation in object oriented code with performance evaluation on network centric terrorist cognizant data on a CDROM with electronic battlefield WTPPP Protocol Software and National Missile Defense(NMD) Protocol Software selected for CTPA Protocol Software Attachment. The complex star overlay CCPDP Protocol geometric software structure represents twenty critical points for each of the eight Catastrophe theory equations. The critical points will be for both the Catastrophic Lagrangian equation objective functions and the constraint functions. Security multilayering of the geometric software structures enables protocol architecture security layering and multileveling for prediction of layers of critical points for the objective function and the constraint function in the Lagrangian equation eight sided geometric star ostic approach. This is the same type of analysis that would occur at sensor holographic image determination[15] for the space based airborne lasers and their sensitive deformable shields and the ground based homeland security objects.

Artificial Intelligence equipment maintenance and support described as a prognostic approach to Homeland Security and Theater Missile Defense requires geometric software structure equation assembly for the CCPDP Protocol to attain computational processing speed. illustrates the eight sided star whose facets each represent one of the eight catastrophe theory equations and their possible eight possible critical points. Homeland Defense and Security systems have used overlays and underlays to obtain additional layers of security. The overlay of a flower[7] to the eight sided star enables the attainment of ten critical points. The twenty critical pointare obtained for each of the eight equations, whether they are the eight Catastrophe theory equations or other equations representing the the image from the sensor more accruately.

## XII. HOMELAND DEFENSE AND SECURITY GNNO PROTOCOL GEOMETRIC STRUCTURE

The UIS GNNO Protocol is converted to the HDSUIS GNNO Protocol by the addition of the Catastrophe Theory Partition Attachment(CTPA) and Catastrophic Critical Point Detection Prediction(CCPDP) Software Protocols to the GNNO Admissibility Cube. Refer to Figure 1 for the Geometric Software structure of the HDSUIS GNNO Protocol containing the UIS Enhancement, Alteration and Attachment Conversions within the Admissibility Cube and the Outer Construction Cube retaining the Message Path Delay with the timing chart and IDE Alterations.

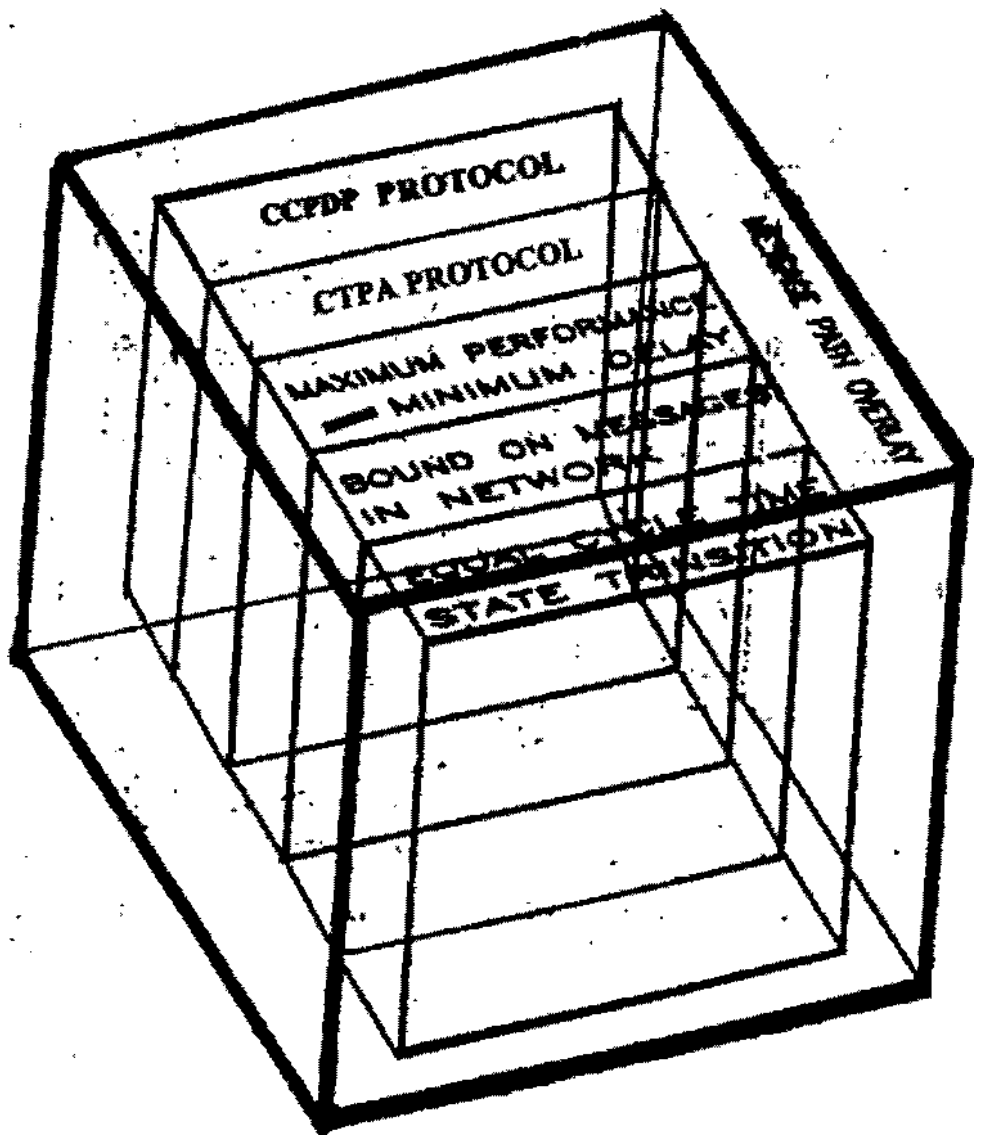

.**Figure 1 HOMELAND DEFENSE AND SECURITY GEOMETRIC NEURAL NETWORK OVERLAY (HDS GNNO) PROTOCOL**

## XIII. HDSUIS OVERALL GEOMETRIC SOFTWARE STRUCTURES

Figure 2 describes the geometric software structure complete transit from the sensors of the CTPA GNNO Protocol and the CCPDP Protocol through the HDSUIS Geometric Software Structure Protocol, and the hardware linkage of the CTPA Protocol Communication Parity Geometric Software Structure between the HDSUIS Geometric Software Structure and the GIG. The order of this transit is crucial to ensure optimal GIG simulation facility determination of actions against terrorist activities detected by the sensors and the timing chart and IDE optimization equation delay predictions.

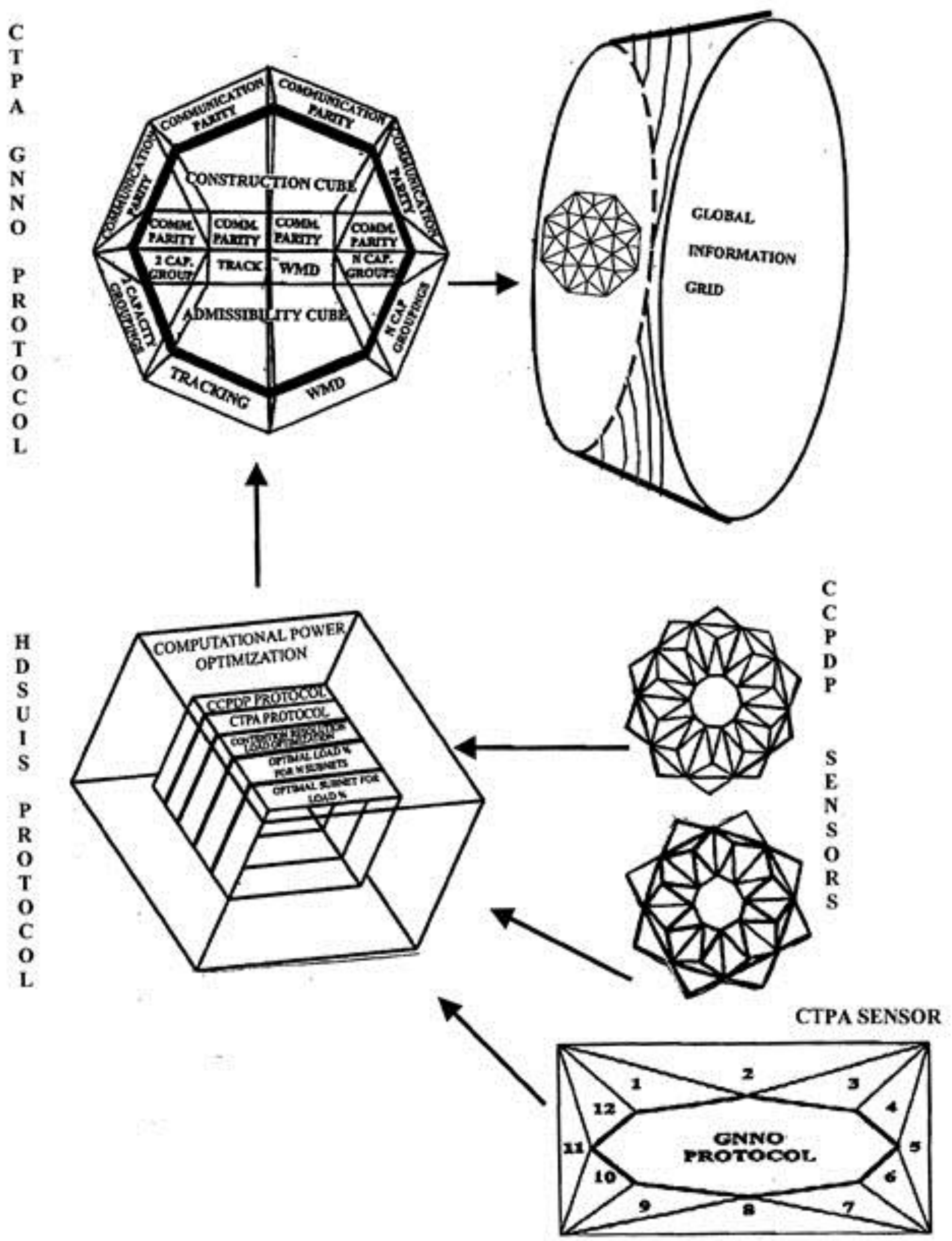

**Figure 2 HOMELAND DEFENSE AND SECURITY UNIVERSAL INTERFACE SOFTWARE(HDSUIS) PROTOCOL including GIG CTPA GNNO PROTOCOL AND CTPA and CCPDP Sensors.**